\documentclass[prd,showpacs,showkeys,nofootinbib,floatfix,eqsecnum,
               fleqn,preprint,12pt,tightenlines]{revtex4-1} %for preprint

\usepackage{amsmath,amssymb,revsymb,graphicx,dcolumn}

\newcommand{\version}{v3} %%true version v2.991
\newcommand{\beq}{\begin{equation}}
\newcommand{\eeq}{\end{equation}}
\newcommand{\beqa}{\begin{eqnarray}}
\newcommand{\eeqa}{\end{eqnarray}}
\newcommand{\bsubeqs}{\begin{subequations}}
\newcommand{\esubeqs}{\end{subequations}}

\begin{document}

\begin{widetext}
\noindent arXiv:2008.11699 \hfill KA--TP--10--2020\;(\version)
%
%\noindent journal  \hfill    arXiv:2008.11699\;(\version)
%
%\noindent \hfill KA--TP--10--2020\;(\version)
%
\newline\vspace*{3mm}
\end{widetext}

\title{IIB matrix model: Extracting the spacetime metric}

\author{\vspace*{5mm} F.R. Klinkhamer}
\email{frans.klinkhamer@kit.edu}
\affiliation{Institute for Theoretical Physics,
Karlsruhe Institute of Technology (KIT),\\
76128 Karlsruhe, Germany\\}

\begin{abstract}
\vspace*{5mm}\noindent
The large-$N$ master field of the Lorentzian IIB matrix model
is of course not known, but we can assume that we already have it 
and investigate how the emerging spacetime metric could be extracted.
We show that, in principle, it is possible to obtain both the Minkowski 
metric and the spatially flat Robertson--Walker metric.
\end{abstract}

%\pacs{04.20.Cv, 98.80.Bp, 98.80.Jk}
%\keywords{general relativity, big bang theory,
%          mathematical and relativistic aspects of cosmology}

%%98.80.Bp Origin and formation of the Universe
%%11.25.-w Strings and branes
%%11.25.Yb M theory


\maketitle

%%\newpage%%tmp
\section{Introduction}
\label{sec:Intro}

The ten-dimensional IIB matrix 
model~\cite{IKKT-1997,Aoki-etal-review-1999} in the Lorentzian version
has provided some hints that a classical spacetime may
emerge with three ``large'' spatial dimensions and six ``small''
spatial dimensions~\cite{KimNishimuraTsuchiya2012,NishimuraTsuchiya2019}.
Still, the conceptual origin of such a classical spacetime
has not been addressed satisfactorily.

Recently, it has been suggested that the large-$N$ master
field~\cite{Witten1979} may play a crucial role for the emergence 
of a classical spacetime manifold from the IIB matrix model
(see App.~B in the preprint version~\cite{Klinkhamer2019-emergence-v6} 
of Ref.~\cite{Klinkhamer2019-emergence-PTEP}).
This suggestion has now been clarified in a triple of
follow-up papers.

The first paper~\cite{Klinkhamer2020-emergent-spacetime-master-field} 
gives a general discussion
of how the classical spacetime points can be extracted from
the bosonic master field of the IIB matrix model 
and how the corresponding classical spacetime metric is obtained. 

The second paper~\cite{Klinkhamer:2020-extracting-points} 
presents an explicit calculation
of how the spacetime points appear in the bosonic master field,
under the assumption that master field has been calculated exactly
or that a reliable approximation of the master field has
been found.

The current paper, the third in the series, aims to present
a direct calculation of the spacetime metric, under the assumption 
that the master field (or an approximation of it) is at hand.
In fact, we will try to see what would be required from the  
master field, in order to obtain the Robertson--Walker metric.

All calculations of this paper are analytic and   
\textsc{Mathematica} 5.0~\cite{Wolfram1991} is used. 

%%\newpage%%tmp
\section{Emergent spacetime metric}
\label{sec:Emergent spacetime metric}

Adapting Eq.~(4.16) of Ref.~\cite{Aoki-etal-review-1999}
to our master-field approach, we obtained in
Ref.~\cite{Klinkhamer2020-emergent-spacetime-master-field}
the following expression for the emergent inverse metric:
\beq \label{eq:emergent-inverse-metric-with-rho-averaged}
g^{\mu\nu}(x) \sim
\int_{\mathbb{R}^{D}} d^{D}y\;
\langle\langle\, \rho(y)  \,\rangle\rangle
\; (y-x)^{\mu}\,(y-x)^{\nu}\;f(y-x)\;r(x,\,y)\,,
\eeq
with continuous spacetime coordinates $x^\mu$ having the
dimension of length
and spacetime dimension $D=10$ for the original model.
The meaning of the average $\langle\langle\, \ldots \,\rangle\rangle$
will be discussed later.
We refer to Refs.~\cite{Klinkhamer2020-emergent-spacetime-master-field,%
Klinkhamer:2020-extracting-points} for the details on how
the discrete spacetime points $\widehat{x}^{\,\mu}_{k}$, with
index $k\in \{1,\, \ldots \,,K\}$, are extracted
from the bosonic master field $\widehat{\underline{A}}^{\,\mu}$,
which corresponds to ten $N\times N$ 
traceless Hermitian matrices for
$N=K n$, with positive integers $K$ and $n$.
The limit $K \to \infty$ entails the limit $N \to \infty$,
as long as $n$ stays constant or increases.

The quantities  entering the above integral
are, first, the density function
\beq \label{eq:rho-def}
\rho(y) \;\equiv \;
\sum_{k=1}^{K}\;\delta^{(10)} \big(y- \widehat{x}_{k}\big)
\eeq
for the emergent spacetime points $\widehat{x}^{\,\mu}_{k}$
as obtained in Refs.~\cite{Klinkhamer2020-emergent-spacetime-master-field,%
Klinkhamer:2020-extracting-points}
and, second, the dimensionless density correlation function $r(x,\,y)$
which is defined by
\beq \label{eq:r-def}
\langle\langle\,\rho(x)\,\rho(y) \,\rangle\rangle \;\equiv \;
\langle\langle\, \rho(x)\,\rangle\rangle\; 
\langle\langle\,\rho(y) \,\rangle\rangle\;r(x,\,y)\,.
\eeq
In addition, there is
a sufficiently localized symmetric function $f(y-x)$,
which appears in the effective action of a low-energy scalar
degree of freedom $\phi$ ``propagating'' over the discrete
spacetime points $\widehat{x}^{\,\mu}_{k}$; see
Refs.~\cite{Aoki-etal-review-1999,%
Klinkhamer2020-emergent-spacetime-master-field} for further details.
As this function $f(x)=f(x^{0},\,  x^{1},\, \ldots \,,\,x^9)$
has the dimension of $1/(\text{length})^{2}$,
the inverse metric $g^{\mu\nu}(x)$
from \eqref{eq:emergent-inverse-metric-with-rho-averaged}
is seen to be dimensionless.
The metric $g_{\mu\nu}$ is obtained as the matrix inverse
of $g^{\mu\nu}$.

The average $\langle\langle\, \ldots  \,\rangle\rangle$
entering \eqref{eq:emergent-inverse-metric-with-rho-averaged}
and \eqref{eq:r-def} corresponds, for the extraction 
procedure of the discrete spacetime points $\widehat{x}^{\,\mu}_{k}$ 
from Refs.~\cite{Klinkhamer2020-emergent-spacetime-master-field,%
Klinkhamer:2020-extracting-points},
to averaging over different block sizes $n$
and block positions along the diagonal in the master field.
But it is not really necessary to do this additional averaging
in the integrand of \eqref{eq:emergent-inverse-metric-with-rho-averaged},
as that is already taken care of by the limit $N\to\infty$,
with appropriate block dimension $n \gtrsim \Delta N$
for width $\Delta N$ of the band-diagonal master-field matrices.
In the following, we will just use $\rho(y)$
in the integrand of \eqref{eq:emergent-inverse-metric-with-rho-averaged},
so that we have, for the emergent inverse metric,
\beq \label{eq:emergent-inverse-metric}
g^{\mu\nu}(x) \sim
\int_{\mathbb{R}^{D}} d^{D}y\;
\rho(y) \; (y-x)^{\mu}\,(y-x)^{\nu}\;f(y-x)\;r(x,\,y)\,.
\eeq

The goal of the current paper is to investigate the
integral \eqref{eq:emergent-inverse-metric} 
and to determine what would be
required of the unknown functions $\rho$, $f$, and $r$
[all three tracing back to the IIB-matrix-model master field],
so that the integral gives, in particular, 
the Robertson--Walker inverse metric.

%%\newpage%%tmp
\section{Question and setup}
\label{sec:Question-and-setup}

In this article, we address the following concrete question:
is it at all possible to get the Minkowski inverse metric
and the spatially flat Robertson--Walker inverse metric
from the expression \eqref{eq:emergent-inverse-metric}?

We restrict ourselves to four ``large'' spacetime dimensions, 
setting \beq \label{eq:D-equals-4}
D=4
\eeq
in \eqref{eq:emergent-inverse-metric}, and also
use length units that normalize the IIB-matrix-model length scale $\ell$
as introduced in Ref.~\cite{Klinkhamer2020-emergent-spacetime-master-field},
\beq \label{eq:ell-equals-1}
\ell=1\,.
\eeq
Next, define
\bsubeqs\label{eq:}
\beqa
r(x,\,y) &\equiv& \widetilde{r}(y-x)\;\overline{r}(x,\,y)\,,
\\[2mm]
h(y-x)  &\equiv& f(y-x)\;\widetilde{r}(y-x)\,,
\eeqa
\esubeqs
where
$\overline{r}(x,\,y)$
has a more complicated (but still symmetric) dependence on $x$ and $y$
than just the combination $x-y$ [a trivial example would be
$\overline{r}(x,\,y)=(x^{0})^{2}+(y^{0})^{2}$\,].

Now, change the integration variables $y^\mu$
in the integral \eqref{eq:emergent-inverse-metric}
to $z^\mu\equiv y^\mu-x^\mu$ and get
\beq \label{eq:emergent-inverse-metric-d4z-integral}
g^{\mu\nu}(x) \sim
\int_{-\infty}^{\infty} d z^{0}\;
\int_{-\infty}^{\infty} d z^{1}\;
\int_{-\infty}^{\infty} d z^{2}\;
\int_{-\infty}^{\infty} d z^{3}\;
\rho(z+x) \; z^{\mu}\,z^{\nu}\;h(z)\;\overline{r}(x,\,z+x)\,.
\eeq
Observe that $x^{\mu}$ enters the right-hand side only via the 
density function $\rho$ and the correlation function $\overline{r}$.
This observation plays a crucial role for obtaining the
constant Minkowski inverse metric, as will become clear
in Sec.~\ref{sec:Minkowski-metric}.

Finally, recall that the ten-dimensional Lorentzian IIB
matrix model has coupling constants $\widetilde{\eta}_{KL}$
(with indices $K,\,L\in \{0,\, 1,\, \ldots\, ,\,9\}$), which
give the usual components $\widetilde{\eta}_{\mu\nu}$ for the 
four-dimensional case:
\beq \label{eq:etatilde-munu-4D}
\widetilde{\eta}_{\mu\nu} =
\begin{cases}
 -1 \,,   &  \;\;\text{for}\;\;\mu=\nu=0 \,,
 \\[2mm]
 +1 \,,   &  \;\;\text{for}\;\;\mu=\nu=m\in \{1,\, 2,\, 3\} \,,
 \\[2mm]
 0 \,,   &  \;\;\text{otherwise} \,.
\end{cases}
\eeq

%%\newpage%%tmp
\section{Minkowski metric}
\label{sec:Minkowski-metric}

There are several ways to obtain the Minkowski inverse 
metric from \eqref{eq:emergent-inverse-metric-d4z-integral}.
One way has been discussed in App.~B   %%XXX
of Ref.~\cite{Klinkhamer2020-emergent-spacetime-master-field}.
But perhaps the simplest recipe is to take 
the following \textit{Ansatz}:   
\bsubeqs\label{eq:rho-rbar-h-Ansaetze}
\beqa
\label{eq:rho-Ansatz}
\rho(z+x) &=& 1\,,
\\[2mm]
\label{eq:rbar-Ansatz}
\overline{r}(x,\,z+x)&=& 1\,,
\\[2mm]
\label{eq:h-Ansatz}
h(z) &=&
\xi\,\exp\Big[-(z^{0})^{2}-(z^{1})^{2}-(z^{2})^{2}-(z^{3})^{2}\, \Big]\;
\Big(
\widetilde{\eta}_{00}\,\left[\,\zeta\,(z^{0})^{2}-1\,\right] 
\nonumber\\[1mm]
&&
+\widetilde{\eta}_{11}\,\left[\,\zeta\,(z^{1})^{2}-1\,\right]
+\widetilde{\eta}_{22}\,\left[\,\zeta\,(z^{2})^{2}-1\,\right]
+\widetilde{\eta}_{33}\,\left[\,\zeta\,(z^{3})^{2}-1\,\right]
\Big)\,,
\eeqa
\esubeqs
where the exponential function in \eqref{eq:h-Ansatz}
provides a symmetric cutoff on the integrals 
of \eqref{eq:emergent-inverse-metric-d4z-integral}
and the IIB-matrix-model coupling constants $\widetilde{\eta}_{\mu\nu}$
are given by \eqref{eq:etatilde-munu-4D}.
The expression \eqref{eq:h-Ansatz} could be simplified somewhat,
but it is more instructive to keep the various contributions 
$\big[\zeta\,(z^\mu)^{2}-1\big]$ separate.  
In fact, this $\zeta$ factor in square brackets gives,  
for the special value $\zeta=2$, the following definite
integrals:
\beq
\label{eq:I-n-integrals}
I_{n} \equiv
\frac{1}{\sqrt{\pi}}\;\int_{-\infty}^{\infty} d z\;
z^n\;\big[\,2\,z^{2}-1\,\big]\;e^{-z^2}
=
\begin{cases}
 0 \,,   &  \;\;\text{for}\;\;n=0,\,1 \,,
 \\[2mm]
 1 \,,   &  \;\;\text{for}\;\;n=2 \,,
\end{cases}
\eeq
which are an essential ingredient of our \textit{Ansatz}.

With the resulting expression from
\eqref{eq:emergent-inverse-metric-d4z-integral},
\beq \label{eq:emergent-inverse-metric-Mink}
g^{\mu\nu}(x) \sim
\int_{-\infty}^{\infty} d z^{0}\;
\int_{-\infty}^{\infty} d z^{1}\;
\int_{-\infty}^{\infty} d z^{2}\;
\int_{-\infty}^{\infty} d z^{3}\;
z^{\mu}\,z^{\nu}\;h(z)\,,
\eeq
in terms of the $h$ function \eqref{eq:h-Ansatz},
and the particular numerical values
\bsubeqs\label{eq:zeta-xi}
\beqa 
\zeta&=& 2 \,,
\\[2mm]
\xi&=& 1/\pi^{2} \,,   
\eeqa
\esubeqs
we get
\beq \label{eq:emergent-inverse-metric-Mink-result}
g^{\mu\nu}_\text{\,Mink}(x) \sim
\begin{cases}
 -1 \,,   &  \;\;\text{for}\;\;\mu=\nu=0 \,,
 \\[2mm]
 +1 \,,   &  \;\;\text{for}\;\;\mu=\nu=m\in \{1,\, 2,\, 3\} \,,
 \\[2mm]
 0 \,,   &  \;\;\text{otherwise} \,.
\end{cases}
\eeq
which is the standard inverse metric of
Minkowski spacetime with Cartesian coordinates.
Observe that this inverse metric is independent of $x^\mu$,
because $x^\mu$ has disappeared from the
integral on the right-hand side
of \eqref{eq:emergent-inverse-metric-Mink}.
Essentially the same observation has been made in the 
sentence below Eq.~(4.17) of Ref.~\cite{Aoki-etal-review-1999}.

Note, finally, that we get the Minkowski metric
by taking the matrix inverse of
\eqref{eq:emergent-inverse-metric-Mink-result}.
The resulting covariant 
tensor $g_{\mu\nu}^\text{\,(Mink)}(x)$ has the same components
as the contravariant tensor \eqref{eq:emergent-inverse-metric-Mink-result}.

%%\newpage%%tmp
\section{Robertson--Walker metric}
\label{sec:RW-metric}

Consider the spacetime points with coordinates
\bsubeqs\label{eq:xmu-cosmological}
\beqa 
x^\mu &=& \big(x^{0},\,x^{1},\, x^{2},\, x^{3}\big)\,,
\\[2mm]
x^{0} &=& \widetilde{c}\,t =t\,,
\eeqa
\esubeqs
where $t$ is interpreted as the cosmic-time coordinate 
and $\widetilde{c}$ is set to unity by an appropriate choice 
of the time unit.

We will try to get the spatially flat Robertson--Walker (RW) 
inverse metric by choosing appropriate \textit{Ansatz} 
parameters $r_n$ (see below for their definition), so that
\bsubeqs\label{eq:RW-inverse-metric-goal}
\beqa
g^{00}_\text{\,RW}(x)&\sim&-1\,,
\\[2mm]
g^{mm}_\text{\,RW}(x)&\sim&+1 + c_{2}\,t^{2}+ c_{4}\,t^{4} + \ldots\,,
\eeqa
\esubeqs
for $m\in \{1,\, 2,\, 3\}$ and constants $c_{2}$ and $c_{4}$, 
and with all off-diagonal components vanishing.

We can obtain this result by making one change in the
previous calculation of Sec.~\ref{sec:Minkowski-metric},
namely, by letting the \textit{Ansatz} density function $\rho(y)$ 
be a nontrivial even function of the time-component of the 
coordinate $y^\mu= x^\mu+z^\mu\,$:
\beq \label{eq:rho-Ansatz-for-RW}
\rho(y)=\rho(y^{0}) \ne 1\,.
\eeq
Specifically, we take an even polynomial (distinguished by an overbar)
of order $2 K_0\,$,
\beq \label{eq:rho-Ansatz-for-RW-polynomial}
\overline{\rho}(y^{0})=\sum_{k=0}^{K_0}\,r_{2k}\;\left(y^{0}\right)^{2k}\,,
\eeq
with arbitrary constants $r_{2k}$.
The resulting expression from
\eqref{eq:emergent-inverse-metric-d4z-integral} reads
\beq \label{eq:emergent-inverse-metric-RW}
g^{\mu\nu}(x) \sim
\int_{-\infty}^{\infty} d z^{0}\;
\int_{-\infty}^{\infty} d z^{1}\;
\int_{-\infty}^{\infty} d z^{2}\;
\int_{-\infty}^{\infty} d z^{3}\;
\overline{\rho}(z+x) \; z^{\mu}\,z^{\nu}\;h(z)\;\overline{r}(x,\,z+x)\,,
\eeq
with the $h$ function \eqref{eq:h-Ansatz} 
for numerical values \eqref{eq:zeta-xi} and, 
for the moment, $\overline{r}(x,\,z+x)= 1$.
Observe that, with $r_0=1$ and $r_{2k}=0$  for $k\geq 1$,
we recover the integral \eqref{eq:emergent-inverse-metric-Mink},
from which the Minkowski inverse
metric \eqref{eq:emergent-inverse-metric-Mink-result}
was derived.

Fixing, for definiteness, the $\overline{\rho}$ polynomial
\eqref{eq:rho-Ansatz-for-RW-polynomial} to be tenth order, we obtain 
from \eqref{eq:emergent-inverse-metric-RW} for $\overline{r}=1$, 
with an appropriate choice of \textit{Ansatz} parameters,
\bsubeqs\label{eq:r0-to-r8-from-r10}
\beqa
r_{0} &=& 1\,,
\\[2mm]
r_{2}&=&\frac{7371}{64}\,r_{10}\,,
\\[2mm]
r_{4}&=&-\frac{4599}{16}\,r_{10}\,,
\\[2mm]
r_{6}&=&\frac{11151}{80}\,r_{10}\,,
\\[2mm]
r_{8}&=&-\frac{108}{5}\,r_{10}\,,
\eeqa
\esubeqs
the following inverse metric:
\bsubeqs\label{eq:RW-inverse-metric-from-r10}
\beqa
g^{\mu\nu}(x) &\sim&
\begin{cases}
 -1+\text{O}(t^6) \,,
 &  \;\;\text{for}\;\;\mu=\nu=0 \,,
 \\[2mm]
 +1+c_{2}\,t^{2}+\text{O}(t^{4}) \,,
 &  \;\;\text{for}\;\;\mu=\nu=m\in \{1,\, 2,\, 3\} \,,
 \\[2mm]
 0 \,,
 &  \;\;\text{otherwise} \,.
\end{cases}
\\[2mm]
c_{2} &=& -\frac{567}{8}\,r_{10}\,.
\eeqa
\esubeqs

In this way, we can get \emph{any} Taylor coefficient $C_{2}$
(denoted by an upper-case symbol)
for the space-space components of the RW inverse metric
\eqref{eq:RW-inverse-metric-from-r10}
by choosing an \emph{appropriate} input value for $r_{10}$.
Three technical remarks are in order:
\begin{enumerate}
  \item
The off-diagonal components of
\eqref{eq:RW-inverse-metric-from-r10} vanish because of
an extra spatial factor $z^m$ in the integrand
of \eqref{eq:emergent-inverse-metric-RW} for $\overline{r}=1$, 
where the rest of the integral is symmetric in $z^m$.
  \item
We can focus on the $2\times 2$ block for $\mu,\,\nu \in \{0,\,1\}$,
as the components $g^{22}$ and $g^{33}$ both equal $g^{11}$ and 
all off-diagonal components vanish, $g^{\mu\nu}\sim 0$ for $\mu\ne\nu$.
  \item
The inverse metric \eqref{eq:RW-inverse-metric-from-r10}
is $x^m$-independent because $\overline{\rho}(z+x)$ 
in the integrand 
of \eqref{eq:emergent-inverse-metric-RW}, for $\overline{r}=1$,
only depends on $x^{0}$, according to 
the \textit{Ansatz} \eqref{eq:rho-Ansatz-for-RW-polynomial}.
\end{enumerate}

We can obtain the Taylor coefficients beyond $c_{2}$ by going
to higher orders in the $\overline{\rho}$ polynomial.
For example, by going to twelfth order, we get, 
with appropriate values of the \textit{Ansatz} parameters
$\{ r_{0},\,r_{2},\,r_{4},\,r_{6},\,r_{8}\}$,
the Taylor coefficients $c_{2}=c_{2}(r_{10},\,r_{12})$
and $c_{4}=c_{4}(r_{10},\,r_{12})$.
However, the mapping $(r_{10},\,r_{12}) \to (c_{2},\,c_{4})$
is noninvertible and it is
not possible to get arbitrary values of $(c_{2},\,c_{4})$
from appropriate values of $(r_{10},\,r_{12})$.
In order to be able to obtain arbitrary values
of $(c_{2},\,c_{4})$ from appropriate 
parameters, our \textit{Ansatz} needs to be extended.  

In fact, return to the tenth-order polynomial $\overline{\rho}$,
but now take a nontrivial \textit{Ansatz} for $\overline{r}\,$:
\beq \label{eq:rbar-Ansatz-for-RW}
\overline{r}(x,\,z+x)= 1+ s_{4}\,(x^{0})^{4}\,(z^{0}+x^{0})^{4}   \,,
\eeq
with an arbitrary constant $s_{4}$. 
Inserting the \textit{Ansatz} \eqref{eq:rbar-Ansatz-for-RW} 
in the integrand of \eqref{eq:emergent-inverse-metric-RW}, 
we obtain, with $\overline{\rho}$ parameters
\bsubeqs\label{eq:r0-to-r8-from-r10-and-s4}
\beqa
r_{0} &=& 1\,,
\eeqa
%%\\[2mm]
\beqa
r_{2}&=&
\frac{45\,\big( 152\,s_{4} + 
       63\,r_{10}\,\left[ 416 + 563\,s_{4} \right]  \big) }
     {16\,\left( 640 + 777\,s_{4} \right) }\,,
\eeqa
%%\\[2mm]
\beqa
r_{4}&=&
-\frac{45\,\big( 44\,s_{4} + 
       7\,r_{10}\,\left[ 1168 + 1569\,s_{4} \right]  \big) }
     {2\,\left( 640 + 777\,s_{4} \right) }\,,
\eeqa
%%\\[2mm]
\beqa
r_{6}&=&
\frac{9\,\big( 76\,s_{4} + 
       21\,r_{10}\,\left[ 944 + 1233\,s_{4} \right]  \big) }
     {2\,\left( 640 + 777\,s_{4} \right) } \,,
\eeqa
%%\\[2mm]
\beqa
r_{8}&=&
-\frac{6\,\big( 32\,s_{4} + 
       63\,r_{10}\,\left[ 256 + 323\,s_{4} \right]  \big) }
     {7\,\left( 640 + 777\,s_{4} \right) } \,,
\eeqa
\esubeqs
the following inverse metric:
\bsubeqs\label{eq:RW-inverse-metric-from-r10-and-s4}
\beqa
g^{\mu\nu}(x) &\sim&
\begin{cases}
 -1+\text{O}(t^6) \,,
 &  \;\;\text{for}\;\;\mu=\nu=0 \,,
 \\[2mm]
 +1+c_{2}\,t^{2}+c_{4}\,t^{4}+\text{O}(t^6) \,,
 &  \;\;\text{for}\;\;\mu=\nu=m\in \{1,\, 2,\, 3\} \,,
 \\[2mm]
 0 \,,
 &  \;\;\text{otherwise} \,,
\end{cases}
\\[2mm]
\label{eq:RW-inverse-metric-from-r10-and-s4-coeff-c2}
c_{2} &=&   
-\frac{540\,\big(  
       42\,r_{10}\,\left[ 2 + 3\,s_{4} \right]+s_{4} 
       \big) }     {640 + 777\,s_{4}}\,,
\\[2mm]
\label{eq:RW-inverse-metric-from-r10-and-s4-coeff-c4}
c_{4} &=&
\frac{\big( 945\,r_{10}\,\left[ 80 + 78\,s_{4} -   63\,{s_{4}}^{2} \right] 
            -6\,s_{4}\,\left[ 10 + 273\,s_{4} \right] \big) }{2\,
     \left( 640 + 777\,s_{4} \right) }\,.
\eeqa
\esubeqs
Now, we can get \emph{arbitrary} Taylor coefficients 
$C_{2}$ and $C_{4}$ (denoted by upper-case symbols)
for the space-space components of the RW inverse metric
\eqref{eq:RW-inverse-metric-from-r10-and-s4}
by choosing \emph{appropriate} input values for $r_{10}$ and $s_{4}$.
Specifically, we obtain by
inverting \eqref{eq:RW-inverse-metric-from-r10-and-s4-coeff-c2} 
and \eqref{eq:RW-inverse-metric-from-r10-and-s4-coeff-c4}:
\bsubeqs\label{eq:r10-and-s4-input}
\beqa            %%frk checked
r_{10}^\text{(input)} &=&
\frac{1855\, {C_{2}}^{2} - C_{2}\, \left( 60 - 1554\, C_{4} \right)
+ 1080\, C_{4}}{17010\, \left( 4 - 9\, C_{2} - 8\, C_{4} \right) }\,,
\\[2mm]
s_{4}^\text{(input)} &=&
-\frac{8\, \left( 5\, C_{2} +  6\, C_{4} \right) }{36 - 21\, C_{2}}\,.
\eeqa
\esubeqs

Note, finally, that we obtain the spatially flat Robertson--Walker metric
by taking the matrix inverse of \eqref{eq:RW-inverse-metric-goal}.
In this way, we get the following covariant tensor:
\bsubeqs\label{eq:RW--metric-goal}
\beq
g_{\mu\nu}^\text{\,(RW)}(x) \sim
\begin{cases}
 -1 \,,
 &  \;\;\text{for}\;\;\mu=\nu=0 \,,
 \\[2mm]
 1\big/\big(1+c_{2}\,t^{2}+c_{4}\,t^{4}+ \ldots\big) \,,
 &  \;\;\text{for}\;\;\mu=\nu=m\in \{1,\, 2,\, 3\} \,,
 \\[2mm]
 0 \,,
 &  \;\;\text{otherwise} \,,
\end{cases}
\eeq
\esubeqs
with, for example, explicit coefficients $c_{2}$ and $c_{4}$
from \eqref{eq:RW-inverse-metric-from-r10-and-s4-coeff-c2} 
and \eqref{eq:RW-inverse-metric-from-r10-and-s4-coeff-c4}.

%%\newpage%%tmp
\section{Discussion}
\label{sec:Discussion}

In the present article, we have considered in some detail how
a Robertson--Walker spacetime metric could arise from 
the IIB-matrix-model master field, 
assuming that the master field (or a reliable approximation of it) 
is known. The relevant expression~\cite{Aoki-etal-review-1999,%
Klinkhamer2020-emergent-spacetime-master-field} 
for the emergent spacetime metric is given by a $D$-dimensional  
integral \eqref{eq:emergent-inverse-metric}, 
with functions $\rho$, $f$, and $r$
that follow from the emerged discrete spacetime 
points $\widehat{x}^{\,\mu}_{k}$.

The particular construction we have performed,
for $D=4$, starts from the
Minkowski metric obtained in Sec.~\ref{sec:Minkowski-metric}, 
where we now understand how a constant
($x^\mu$-independent) inverse metric may come about.
By a deformation, we have then found, in Sec.~\ref{sec:RW-metric}, 
the spatially flat ($k=0$) Robertson--Walker 
metric. This immediately raises two questions.

First, is it possible to obtain, in the same way, a Robertson--Walker 
metric with positive ($k=+1$) or negative ($k=-1$) spatial curvature?
\emph{A priori}, we would expect that this is impossible.
Yet, recall that, for example, the $k=1$ Robertson--Walker metric may
not really have an underlying $\mathbb{R}\times S^{3}$ topology 
but can be described
by strong gravitational fields over Minkowski spacetime
with $\mathbb{R}^{4}$ topology 
(see Ref.~\cite{Klinkhamer2012} and references therein).

Second, is it possible to modify the construction 
of Sec.~\ref{sec:RW-metric}, in order to obtain 
the regularized-big-bang  metric~\cite{Klinkhamer2019}?
This question is not quite trivial, 
as the regularized-big-bang \emph{inverse}-metric
component $g^{00}$ diverges at cosmic-time coordinate $t=0$,
where a spacetime defect has replaced the big bang singularity. 
Perhaps it is possible to modify the functions entering the integrand 
of \eqref{eq:emergent-inverse-metric-RW} in such a way that 
the effective cutoff on the integrals for $\mu=\nu=0$
disappears at $x^{0}=t=0$, making the inverse-metric 
component $g^{00}$ diverge at that time slice. 
We hope to address this issue in a future publication.

\section*{Note added}

The paper mentioned in the last sentence of Sec.~\ref{sec:Discussion} 
has meanwhile appeared as Ref.~\cite{Klinkhamer2020-rbb-IIBmm}.

%%%%%%\newpage%%tmp
%\vspace*{-5mm}
%\begin{acknowledgments}
%\vspace*{-5mm}
%It is a pleasure to thank XXXXXXXXXXXXX
%for comments on an earlier preprint version of this article.
%\vspace*{-0mm}
%\end{acknowledgments}

%%%%%%%%%%%%%%\newpage
%\vspace*{-5mm}

\end{document}